\documentclass{aa}
\usepackage{graphics}
\usepackage{natbib}
\begin{document}
\title{Compact Groups in the UZC galaxy sample: II. Connections between 
morphology, luminosity and large-scale density 
\thanks{This research has made use of the NASA/IPAC Extragalactic Database (NED) which is operated by the Jet Propulsion Laboratory, California Institute of Technology, under contract with the National Aeronautics and Space Administration}}  
\author{B.Kelm and P. Focardi} 
\institute{Dipartimento di Astronomia, Universit\`a di Bologna, Via Ranzani 1, 
I-40127 Bologna}
\abstract{
The nature of Compact Groups (CGs) is investigated by comparing the 
luminosities and morphologies of CG galaxies, CG Neighbours and Isolated 
galaxies. CGs turn out to include more early type galaxies than Isolated 
galaxies and fewer low-luminosity galaxies. 
The 33 CGs with a dominant E/S0 and the 30 CGs with a 
dominant spiral have similar LF parameters. 
Spiral dominated CGs have fewer galaxies at high and low 
luminosity in comparison with their Neighbours, while E-S0 dominated CGs 
seem to lack only faint galaxies when compared to their Neighbours. 
Ellipticals which are the dominant galaxy of a CG are also brighter 
than all their Neighbours, while this holds true  for only half of the 
dominant spirals and S0s. 
Relations linking the number of Neighbours of dominant E-S0s to 
the luminosity of E-S0s and to the difference between the first and the 
second ranked CG members do suggest a link between 
the formation of bright early-type galaxies and the presence of a group-like 
potential. No similar relations are found for dominant spirals. 
These tentative results are compatible with the assumption 
that CG dominant Ellipticals are anomalous galaxies whose formation 
might have been a secondary outcome during the process of groups formation.    
}
\titlerunning{CGs, neighbours and isolated galaxies}
\maketitle
\section{Introduction}
Because of their high density and relatively low velocity dispersion,   
Compact Groups (CGs) are predicted to constitute the most probable 
sites for strong galaxy-galaxy interactions and mergers to occur 
\citep{Mamon92a}. If CGs merge into a single galaxy in several Gyr 
\citep{Barnes85,Barnes89,Mamon87}, many remnants, possibly resembling 
bright field ellipticals  \citep{Zablu99,Vikhlinin99,Jones} will be 
produced in less than a Hubble time. 

Indeed, Hickson CGs (HCGs, Hickson 1982, 1997 and references therein) 
do include 
an excess of both, early-type galaxies and luminous sources, and they 
appear as the most likely sites for bright isolated 
Ellipticals to form. 
However, the scarce evidence for strongly interacting 
galaxies \citep{Rubin,Zepf93}, 
the underabundance of peculiar or blue early-type galaxies 
\citep{Zepf91,Zepf91b,Fasano} as well as the non significant 
difference between galaxies in HCGs and 
in other environments as far as the fundamental plane is concerned 
\citep{Rosa} do leave open the question of CG evolution.  
Dynamical N-body simulations \citep{Mamon92b} indicate indeed, that 
many of the global characteristics of HCGs lying in between those 
of field galaxies and those of strongly interacting binaries are consistent 
with the assumption that chance alignments of galaxies in small groups 
are binary-rich.   

Hickson's sample includes CGs spanning a wide range of characteristics, 
and a relevant fraction of HCGs might not be isolated 
\citep{Sulentic87,Williams,Rood,Mamon90,Palumbo}. 
In an attempt to produce a larger and more uniform CG sample,   
Focardi \& Kelm (2002) have automatically selected 291 CGs (UZC-CGs) 
in the Updated Zwicky Catalog (UZC) redshift catalog \citep{Falco}. 
 Contrary to HCGs, which represent surface brightness enhancements, 
UZC-CGs are selected according to compactness criteria (surface number 
density enhancement) and are thus less biased to compact groups of very 
luminous galaxies. 

Focardi \& Kelm have shown that, at variance with 
galaxies lacking  close companions, UZC-CGs are typically non isolated 
systems  displaying an excess of gas-poor galaxies. 
Further they suggest CGs selected according to compactness 
criteria to constitute an intrinsically non-homogeneous sample, 
including two distinct classes of systems which segregate according to 
morphology and large scale environment.      

One class of CGs displays small velocity dispersion, a high fraction of 
emission-line galaxies and is typically located in sparse environments. 
This class constitute a possibly genuine sample of isolated field-CGs 
\citep{Ribeiro} probably rich in AGNs \citep{deCarvalho99,Coziol,Kelm03c}, 
but its high content in gas-rich galaxies suggests it might be contaminated by 
non-bound accordant redshift projections. 
\begin{table*}
\begin{center}
\caption[] {Morphological content of the samples.}
\begin{tabular}{|l||rrrrrrr|}
\noalign{\smallskip}
\hline
\hline
\noalign{\smallskip}
sample &N$_{tot}$ & N$_E$ & N$_{S0}$ & N$_{Sa-Sb}$ &N$_{Sc-Sd}$ & N$_{Spir}$& early-type fraction\cr 
\hline
\noalign{\smallskip}
CGs           & 220  & 25 & 44  &  40 & 39    &30 & 40\%\\
Neighbours & 278  & 15 &46   &  52 & 65    &46 & 27\% \\
Isolated      & 386  &  9 &36   &  62 & 123    &44 & 16\%\cr 
\noalign{\smallskip}
\hline
\hline
\end{tabular}
\end{center}
\end{table*}

The other class of CGs, associated with embedded systems, displays  
large velocity dispersion and includes high fractions of absorption-line 
galaxies. Since CGs presenting a passive population are embedded in 
fairly dense environments, a doubt emerges on whether these galaxies 
can be considered genuine compact group members rather than collapsing 
cores within large groups \citep{Governato96} or transient projections 
within loose groups \citep{Mamon86,Diaferio00}, clusters \citep{Walke89} 
or long cosmological filaments seen end-on \citep{Hernquist95}. 
The question is clearly relevant, for the non-reality of CGs would 
imply that no (or very few) galaxy systems exist on scale intermediate 
between the scale of galaxies and the scale of groups. 

If CGs are real physical structures their local high galaxy density 
is expected to affect galaxy properties. It can simply act as a particular 
initial condition, or influence subsequent galaxy evolution or both. 
In any event, the dominant mechanism acting in CGs is expected to 
significantly alter the luminosity and the morphology of galaxies 
within CGs \citep{Carlberg01,Helsdon03a}. 
To determine which characteristics are intrinsic 
of the galaxies rather than of CGs, we compare galaxies in CGs 
with a sample of Isolated galaxies. 
Similarly, to test contamination of the CG sample by transient configuration 
within loose groups, we compare CGs with their large scale 
Neighbours. 

We will also evaluate the role of a dense environment 
on the evolution of spirals and early-type galaxies separately, 
and discuss differences between CGs whose dominant member is a 
spiral and a E-S0 respectively. This allows addressing the 
controversial issue of the true nature of spiral rich groups, 
and to comment on the lack of observed extended X-ray emission in spiral-only 
groups \citep{Ebeling94,Ponman96,Mulchaey96,deCarvalho99,Mulchaey03}. 
Relative to previous studies addressing similar topics our analysis 
has the great advantage of comparing samples selected within 
the same flux-limited catalogue \citep{Falco} and the same 
redshift range.  
A Hubble constant of $H_0$=\,100\,$h^{-1}$\,km\,s$^{-1}$\,Mpc$^{-1}$ 
is used throughout.  
\section{Isolated Galaxies, CGs and Neighbours: the samples}
All samples are automatically 
selected in the UZC, which is 96\% complete for northern galaxies with 
$m_B$$\leq$$15.5$. Magnitudes have been estimated by eye by Zwicky. 
Bothun \& Cornell (1990) estimate the photometric accuracy to be 
$\approx$ 0.3 mag.   

A galaxy is defined Isolated when no companion galaxies 
are found within a region of 1\,$h^{-1}$\,Mpc projected radius and 
1000 km\,s$^{-1}$ from it. 
It is worth stressing that our sample of Isolated galaxies 
is based on a nearly complete catalogue whose limiting magnitude is 
$\approx$1 magnitude fainter than those used in previous 
investigations \citep{Colbert01,Helsdon01}. 
  
A UZC-CG is defined as a system including at 
least 3 galaxies within a region of 200\,$h^{-1}$\,kpc projected 
radius and radial velocity within 1000 km\,s$^{-1}$ from its center 
\citep{FK}. Cluster 
subclumps have been excluded from the UZC-CG sample rejecting all CGs
found within 1.5\,$h^{-1}$\,Mpc from ACO clusters.   
Samples of Isolated galaxies and CGs at high galactic latitude 
($|$b$^{II}$$|$ $\ge$ 30$\degr$) and $\delta$$\geq$-2$\degr$ 
are here analysed, covering a solid angle of $\sim$ $\pi$ sr.  
 We do not consider systems at lower galactic latitude because  
galactic extinction would bias the samples, artificially 
enlarging the fraction of Isolated galaxies.

We investigate systems in the range of radial velocities between 
2500 and 5500 km\,s$^{-1}$. The lower cut is 
adopted to avoid misidentification of group members whose Hubble flows 
are strongly biased by peculiar velocities. 
The upper limit is adopted to investigate a galaxy sample which is 
relatively complete (in flux and volume) also well below 
$M_*$. At a distance limit of 5500 km\,s$^{-1}$, the UZC extends to  
$M_B$=$-18.2$ + 5\,log\,{\it h} which is  roughly 1.5 magnitude fainter 
than $M_*$, the knee of the Schechter (1976) luminosity function. 

The Isolated galaxy sample includes 386 galaxies, the CG sample includes 
220 galaxies lying in 69 UZC-CGs. 
Most CG galaxies (177=77\%) are in Triplets (Ts). The excess of Ts 
among CGs is partially induced by having rejected subsystems of ACO clusters 
from the sample, as well as non-symmetric systems (CGs whose galaxies 
have no univocal membership assignment on a 200\,$h^{-1}$\,kpc projected 
radius scale). In the range of radial 
velocities  between 2500 and 5500 km\,s$^{-1}$, we have rejected 
43 non-symmetric groups, 26 including 4 or more members.

Neighbours are galaxies at distances between 0.2 $h^{-1}$ and
 1 $h^{-1}$ Mpc and at radial velocity within 
1000 km\,s$^{-1}$ from the CG centers. The Neighbour sample includes 
278 galaxies.  Naked CGs, i.e. with 0 Neighbours are  rare (3 out of 69 CGs).  

Morphological classification is available (NED) for most galaxies 
in the samples (82\% in CGs, 86\% in Neighbours, 71\% in the 
Isolated galaxy sample). Table 1 lists the number of galaxies per 
morphological class, in the 3 samples. N$_{Spir}$ indicates spirals which 
cannot be attributed to Sa-Sb or Sc-Sd classes. 
The last column lists the early type fraction in the samples,   
normalized to the number of galaxies for which morphological classification 
is available.  
\section{Isolated galaxies and galaxies in CGs: do they constitute 
two different populations?}
\begin{figure}
\resizebox{\hsize}{!}{\includegraphics{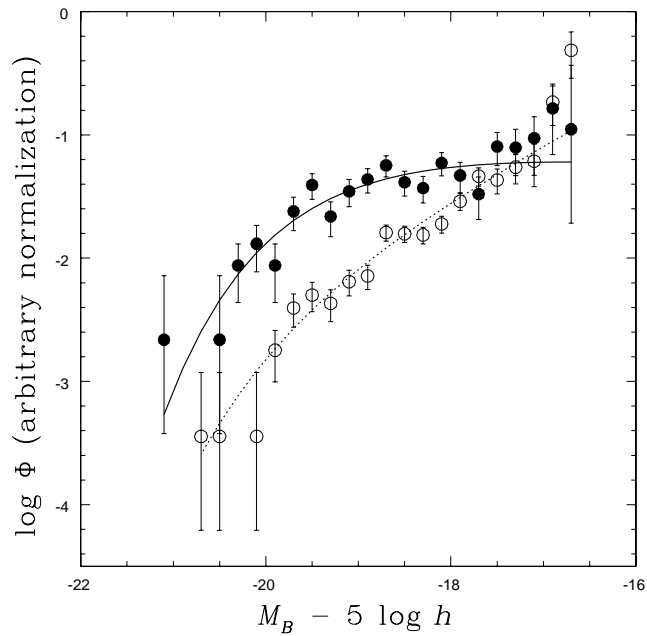}}
\hfill
\caption[]
{Luminosity functions for CG galaxies (filled circles) and 
Isolated galaxies (open circles). 
The points are computed with a modified version of the C-method. 
1$\sigma$ Poissonian uncertainties are plotted.      
Best fits are shown, derived with the STY method in the case of a single 
Schechter function. The relative normalization of the CG 
versus Isolated samples is arbitrary}. 

\end{figure} 
\begin{table*}
\begin{center}
\caption[] {Schechter best fit parameters for CGs, Neighbours and for the 
Isolated galaxy sample.}
\begin{tabular}{|l||rrc|}
\noalign{\smallskip}
\hline
\hline
\noalign{\smallskip}
sample & N$_{tot}$ & $M_{*}$ & $\alpha$ \cr 
\hline
\noalign{\smallskip}
CGs (all)         & 220 & --19.3  & --0.93 \\
CGs (E-S0)        &  69 & --18.7  & +0.56 \\
CGs (Spirals)     & 110 & --19.1  & --0.79 \\
Iso (all)         & 386 & --19.7  & --2.08 \\
Iso (E-S0)        &  45 & --18.2  & +0.38 \\
Iso (Spirals)     & 229 & --19.5  & --1.65 \\
Neigh (all)       & 278 & --20.1  & --1.82 \\
Neigh (E-S0s)     &  61 & --19.0  & --0.49 \\
Neigh (Spirals)   & 176 & --20.1  & --1.84 \cr 
\noalign{\smallskip}
\hline
\hline
\end{tabular}
\end{center}
\end{table*}
Figure 1 shows luminosity functions for the 386 
Isolated (open circles) and the 220 CG (filled circles) galaxies.  
The points are computed with a modified version \citep{Zucca97} 
of the C-method \citep{Lynden71}. Schechter function parameters 
(see Table 2) are derived  with the STY method \citep{Sandage79}. 
Error bars represent 1$\sigma$ Poissonian uncertainties.      
The sample of Isolated galaxies display brighter M$_*$ than CG galaxies, 
however, one should consider that in the isolated sample there are only 
3 galaxies more luminous than M=$-20$, while in the CG sample 12 such galaxies 
are found.     
Table 2 also indicates that the number of dwarfs in CGs is considerably 
smaller than the number of dwarfs in the Isolated galaxy sample.
The fact that CGs are poor in faint galaxies  
could imply that CGs are formed without low luminosity 
galaxies, but it is also consistent with a scenario 
claiming that they have lost  or cannibalized them. 

The value of the faint-end slope $\alpha$ has been a matter of debate for 
HCGs: Mendes de Oliveira \& Hickson (1991) find $\alpha$=$-0.2$, while 
Sulentic \& Rabaca (1994) report $\alpha$=$-1.13$ (and $\alpha$=$-1.69$ for 
the $M$$<$$-18$ galaxy subsample). Zepf et al. (1997) use redshifts of faint 
galaxies around 17 HCGs to replenish the faint end of the HCG luminosity 
function. The authors find $\alpha$=$-1$ and state that HCGs are not 
underabundant of intrinsically faint galaxies. 
Hunsberger et al. (1998), fit the distribution of 39 HCGs 
with 2 Schechter functions and interpret the decreasing 
bright end slope as a deficit of intermediate luminous galaxies.  

While results concerning HCGs are inevitably biased because of the magnitude 
concordance criterion ($\Delta$m $\leq$3) applied to select the groups, 
in the UZC-CG sample the lack of faint galaxies relative to the 
Isolated sample is real, as the redshift range 
and the magnitude depth of the CG and Isolated galaxy samples are the 
same. 
\begin{figure}
\resizebox{\hsize}{!}{\includegraphics{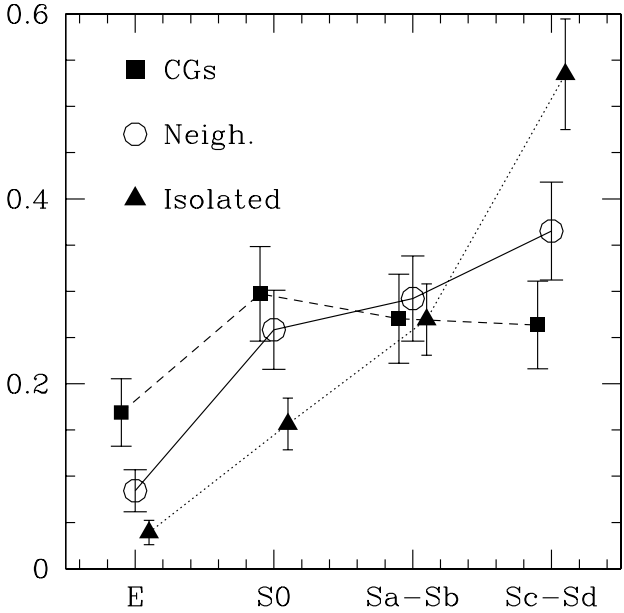}}
\hfill
\caption[]{Morphological distribution of 
CGs, Neighbours and Isolated galaxies. Error bars are multinomial, 
points are slightly shifted for clarity.   
CGs display an excess of ellipticals and a deficiency of 
late spirals compared to Isolated galaxies. 
Neighbours display a morphological distribution intermediate 
between the distributions of Isolated and CG galaxies. 
}
\end{figure}
It has been suggested \citep{Zablu} that  
the deficiency of faint galaxies in CGs might reflect the enhanced 
rate of dynamical friction for luminous (massive) galaxies, bringing high 
luminous galaxies into the compact core of loose groups, and leaving low 
luminosity galaxies outside the loose group cores. 
\section{The influence of local density on morphology and luminosity}
CG galaxies are expected 
to display significant differences in morphological content compared  
to Isolated galaxies. 
Figure 2 shows the morphological content of Isolated galaxies,  
CGs and Neighbours. 
This figure indicates that CGs display a 
significant excess of early-type galaxies, along with a clear deficiency 
of late ($>$ Sb) spirals when compared to Isolated galaxies. 
That CGs contain a relatively higher fraction of elliptical galaxies 
relative to the field has also been found in the SDSS-CG sample \citep{Lee03}. 
Neighbours display a distribution intermediate between CGs and 
Isolated galaxies, suggesting galaxies surrounding CGs are neither  
typical CG members nor typical Isolated galaxies. 
The relative content of early-spirals is the same in the 3 samples 
indicating that the fraction of  Sa-Sb galaxies is a poor tracer of 
environmental density. 
\begin{figure}
\resizebox{\hsize}{!}{\includegraphics{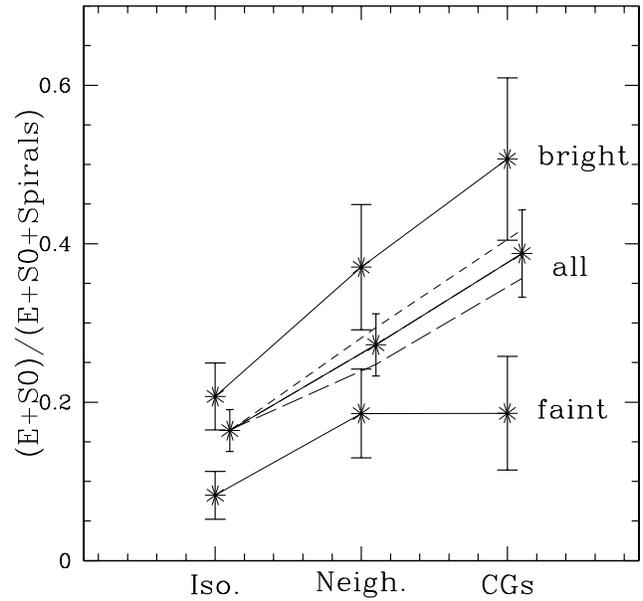}}
\hfill
\caption[]{Global morphology-density relation for Isolated galaxies, 
CGs and Neighbours. The fraction of early-type galaxies (E-S0s) 
normalized to the number of galaxies with assigned morphological type, 
for each of the 3 samples is plotted. Points are shifted for clarity, 
error bars are multinomial.  
Fractions have been computed for the total samples (central) 
and for the faintest and more luminous thirds of each sample. 
The early-type fraction appears to increase  
as a function of both, galaxy density and luminosity.    
}
\end{figure} 

In Fig. 3, the fraction of E-S0s over the total number of galaxies 
with assigned morphological type in each of the 3 samples is plotted.  
As one can reasonably expect from the morphology-density 
relation \citep{Postman84}, the early-type fraction does increase  
with global density, ranging from 16\% in the Isolated galaxy sample to 
39\% in the CG sample. 
The fraction of early-type galaxies we find in UZC-CGs is identical to 
the fraction found in HCGs, when limiting Hickson's sample to groups 
lying between 2500 and 5500 km\,s$^{-1}$.   
Mergers could be responsible for the fact that Neighbours 
display an intermediate global fraction of E/S0s (27\%) between those of 
CGs and Isolated galaxies (see Mamon 2000). 
Alternatively, Fig. 3 could indicate that CGs are typically born in 
environments which are already evolved compared to the low density field, 
suggesting that whatever the enhancement of early-type galaxies in CGs, 
it is partially due to initial conditions. 
\begin{figure}
\resizebox{\hsize}{!}{\includegraphics{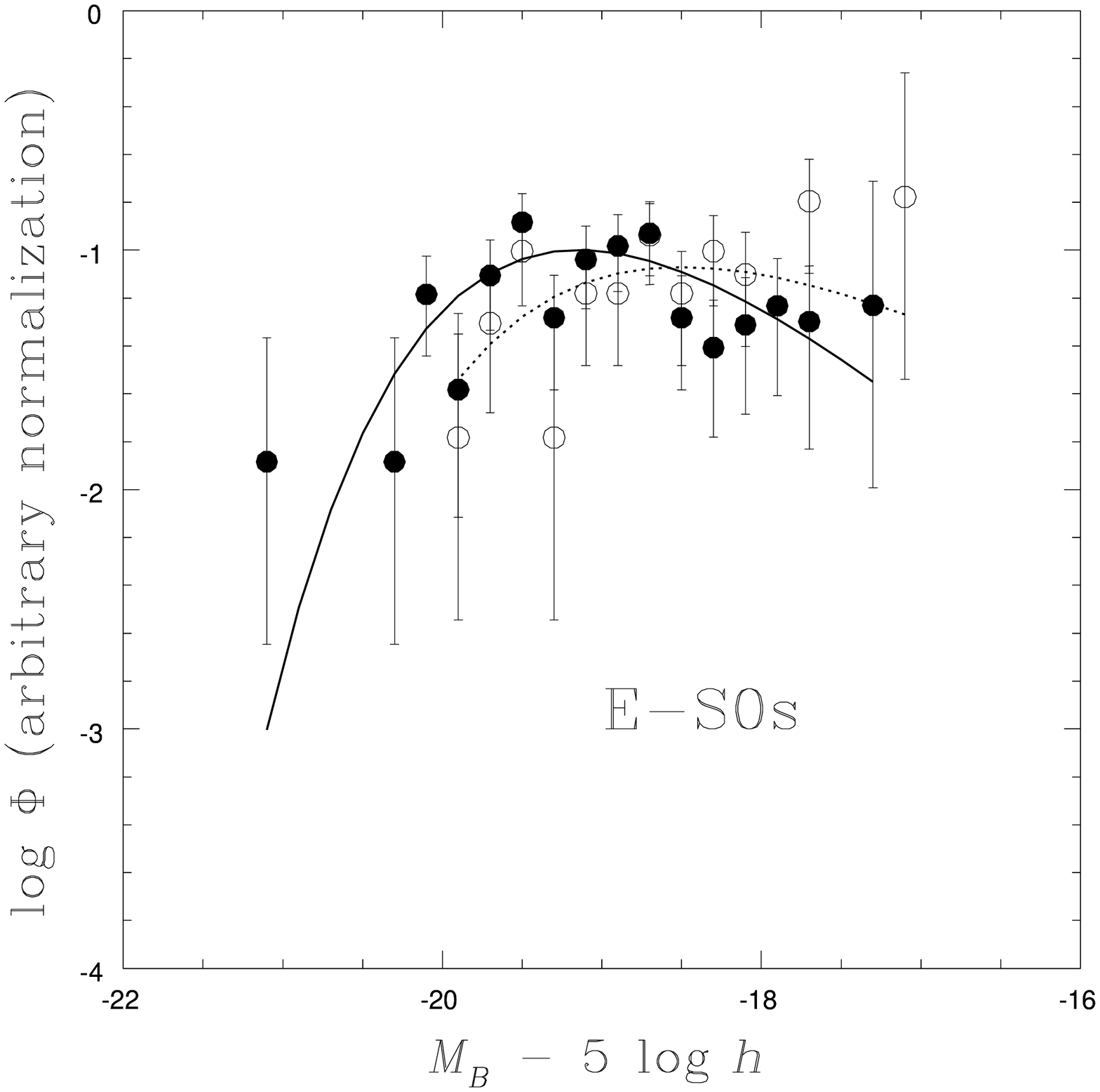}}
\resizebox{\hsize}{!}{\includegraphics{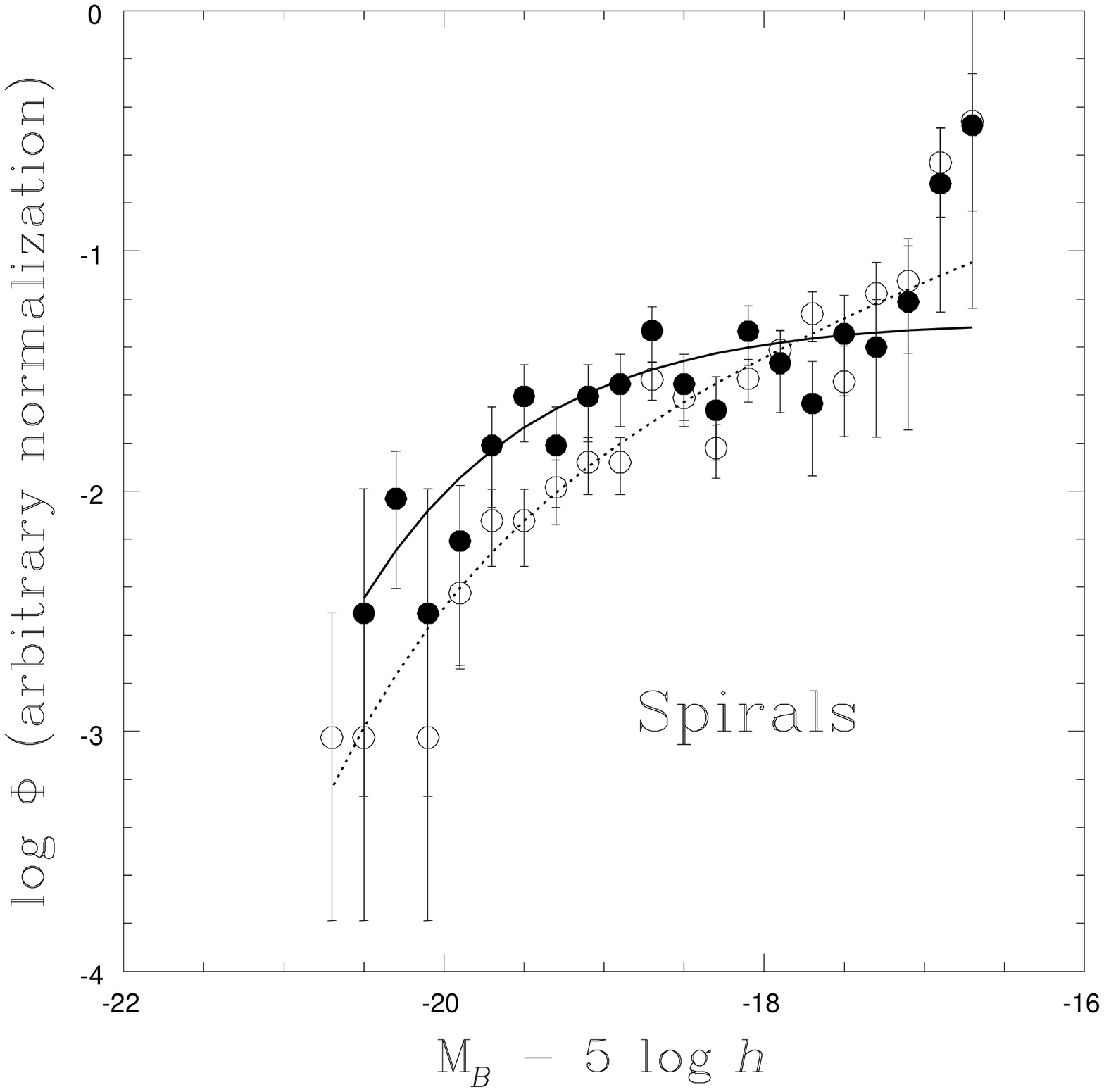}}
\hfill
\caption[]{Luminosity functions for E-S0 and spiral galaxies in the isolated 
galaxy sample (open circles) and in CGs (filled circles). 
The points are computed with a modified version of the C-method. 
1$\sigma$ Poissonian uncertainties are plotted. 
Best fits are shown, derived with the STY method in the case of a single 
Schechter function.The relative normalization of the CG versus Isolated 
samples is arbitrary.
}
\end{figure} 

Because the early type fraction increases in more luminous samples,  
we also show in Fig. 3 the early-type fractions computed for the  
faintest and more luminous thirds of each sample.  
The early-type fraction increases with 
environmental density in the brightest subsample, 
but not in the faintest one. 
Figure 3 indicates that only the morphologies of high luminosity 
CG galaxies are affected, and might suggest that the CG environment 
enhances the luminosity of early-type galaxies only, while leaving the 
luminosity of the spiral population quite unaffected.  
And indeed, while early-type galaxies do constitute only 30\% of the 
whole CG galaxy population (when including also galaxies with no 
morphological type assigned), they 
represent roughly half of the CG dominant members. 
\section{Spirals and E-S0s: type specific luminosity distributions}
The analysis performed so far suggests the CG environment 
is a favourable one for bright E-S0s and could imply that 
the high luminosity of E-S0s is (or has been) triggered by the CG.   
It remains then to be seen whether spirals are triggered similarly:   
to investigate this point, type specific luminosity distributions
of CGs and Isolated galaxies are to be compared.  

Figure 4 shows the luminosity functions for E-S0 and spirals separately 
in the CG and in the Isolated sample. 
In CGs, spirals display steeper faint-end slopes ($\alpha$) than do E-S0s 
(see Table 2 and Fig. 5 for contour plots). The same result is found when 
comparing spirals and E-S0s in the Isolated galaxy sample. 
This confirms, for CGs and for the isolated environment separately, 
the results by Lin et al. (1996), 
Zucca et al. (1997) and Madgwick et al. (2002) all claiming emission line 
galaxies have a steeper faint end slope than galaxies without emission lines. 

The faint-end luminosity functions are steeper for isolated galaxies than 
for CG galaxies in the spiral samples. 
In the E-S0s samples the difference is small, 
suggesting that the lack of dwarf galaxies in CGs might be related to 
their higher fraction of early-type galaxies \citep{deLapparent}. 
The values of the parameter $M_*$ for E-S0s and spirals in the isolated 
sample indicate that bright isolated galaxies are typically spirals.  
In CGs the values of $M_*$ for E-S0s and spirals are roughly similar: 
among the brightest CG galaxies ($M<-20$) 7 E/S0 and 5 spirals are 
found, suggesting that in general luminous galaxies are more likely associated 
to early-type hosts.  
\section{Galaxies in CGs and CG Neighbours}
It has been shown in Fig. 3 that CG Neighbours display a fraction of 
early-type galaxies intermediate between CGs and isolated galaxies. 
It has also been shown that the fraction of luminous galaxies among the 
E-S0s is somewhat larger in CGs than among their Neighbours. 
To test whether CG Neighbours present a deficit of 
bright E-S0s compared to CGs, 
we next compare E-S0s in the Neighbour and CG samples. In Fig. 5 we show 
confidence ellipses (at 1$\sigma$) for the parameters $\alpha$ and M$_*$ 
for CGs (solid), Neighbours (dashed) and Isolated galaxies (dotted).
\begin{figure}
\resizebox{\hsize}{!}{\includegraphics{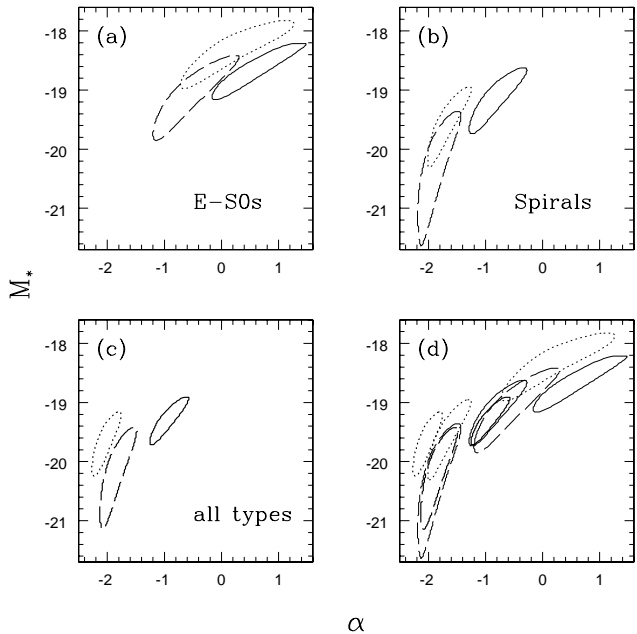}}
\hfill
\caption[]{
Confidence ellipses for the parameters $\alpha$ and $M_*$, 
for CG galaxies (solid), Neighbours (dashed) and Isolated galaxies (dotted). 
1$\sigma$ contours are plotted. Panels {\bf a} and {\bf b} refer to E-S0s and 
spirals subsamples, panel {\bf c} to full samples. Panel {\bf d} shows all 
samples together. 
CGs appear to have fewer low-luminosity galaxies than Neighbours and than 
Isolated galaxies, and no excess of bright galaxies.       
}
\end{figure} 

Figure 5 shows that LF parameters are strongly dependent on 
morphological type, regardless of environment. 
But it also indicates that CGs have fewer low-luminosity E-S0s than their 
Neighbours, and definitely fewer low-luminosity spirals.  
Figure 5 suggests that the lack of faint galaxies in CGs 
is not a characteristic shared by their neighbour galaxies. 
Hence, it appears that a segregation \citep{Zablu} between bright and 
faint galaxies occurs, separating CG galaxies from their neighbours.  

LF contour plots indicate marginal brighter $M_*$ for Neighbour E-S0s  
relative to E-S0s in CGs, and significative brighter $M_*$ for 
Neighbour spirals relative to spirals in CGs. 
No excess of bright E-S0s is associated to CGs 
in terms of their $M_*$ parameter. Along with the deficit in dwarf E-S0s, 
this suggests that CGs include an excess of E-S0s of intermediate luminosity 
relative to E-S0s in their Neighbour sample. 
The excess of E-S0s of intermediate luminosity in CGs appears to 
contradict the finding of Hunsberger et al. (1998) that HCGs show a deficit 
of intermediate luminous galaxies relative to the field. 

The lack of bright spirals in CGs in comparison with 
spirals in the Neighbour sample confirms that dense CGs are not 
a preferred environment for bright spirals. 
\begin{figure}
\resizebox{\hsize}{!}{\includegraphics{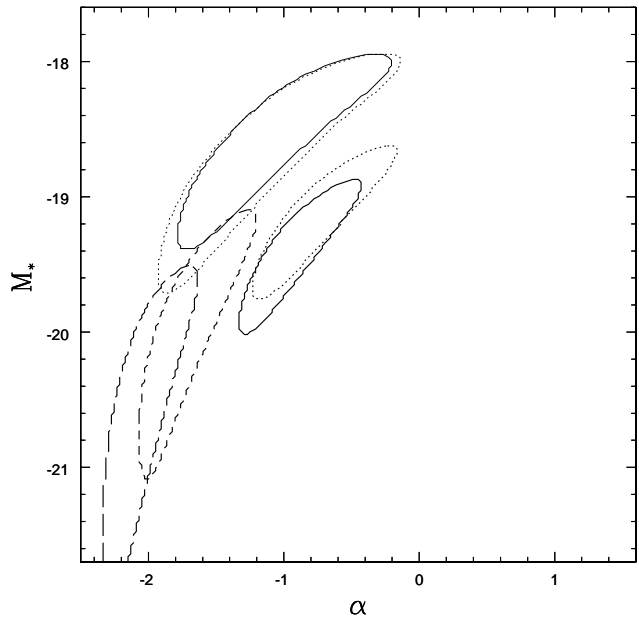}}
\hfill
\caption[]{Confidence ellipses for the parameters $\alpha$ and $M_*$ 
for E-S0 dominated CGs (solid) and Spiral dominated CGs (dotted). 
Heavy contours are for the whole samples, light contours for 
CGs without their dominant members.  
1$\sigma$ contours are plotted.  
Confidence ellipses for neighbours of CGs with a dominant E-S0 
(short dashed) and a dominant Spiral (long dashed) are also shown.   
E-S0 dominated CGs appear to have fewer low-luminosity galaxies than their 
Neighbours. Spiral dominated CGs appear to have significantly fewer 
dwarfs and bright galaxies in comparison with their Neighbours. 
}
\end{figure} 
Differences between late and early type galaxies are also seen 
in loose group samples of bright galaxies \citep{Girardi03} with  
early-type galaxies lying closer to the group center than late type 
galaxies.  

Results drawn from CG type specific analysis deserve some 
caution, however, because real CGs and their Neighbours  
are never E-S0-only (spiral-only) systems. 
\begin{table*}
\begin{center}
\caption[] {Morphological content of E-S0 and Spiral dominated CGs 
and of their Neighbour samples.}
\begin{tabular}{|l||rrrrrr|}
\noalign{\smallskip}
\hline
\hline
\noalign{\smallskip}
sample & N$_E$ & N$_{S0}$ & N$_{Sa-Sb}$ &N$_{Sc-Sd}$ & N$_{Spir}$& early-type fraction\cr 
  & & & & & & with/without dom. \cr 
\hline
\noalign{\smallskip}
E-S0 dom. CGs           & 19 & 30  &  14 & 15   &15 &53\%  /  27\%\\
E-S0 dom. Neigh.        & 7 & 21  &  23 & 29    &20 &  28\%\\
Spiral dom. CGs         & 5  & 12  &  28 & 22   &15 &21\%  /  33\%\\
Spiral dom. Neigh.      & 6 & 23  &  26 & 31    &24 &  26\%\cr 
\noalign{\smallskip}
\hline
\hline
\end{tabular}
\end{center}
\end{table*}
\section{Dominant CG galaxies, non-dominant CG galaxies and Neighbours}
We next investigate whether differences between CG and Neighbour samples 
are equally likely induced by CGs which are dominated by early-type 
galaxies and by spirals. 
In comparison with the global type-dependent luminosity function, 
the analysis of luminosity functions according to the morphological type 
of the dominant member seems more meaningful. 
Dynamical simulations of similar mass merging 
spirals lead to ellipticals for binary mergers and for multiple mergers 
within dense groups \citep{Barnes88,Barnes89}. 
Spiral dominated and E-S0 dominated CGs also display  
a very different behaviour in the X-ray domain: E-S0s dominated CGs are 
likely associated to extended X-ray emission \citep{Mulchaey03} and 
E-S0 dominated CGs might constitute an artificial subsample of 
E-S0 dominated loose groups \citep{Zablu,Helsdon00}. 
At variance with this, it is still uncertain whether spiral dominated  
(or spiral rich) CGs do exist presenting diffuse X-ray emission 
\citep{Burstein02,Mulchaey03} as AGNs seem to contaminate the X-ray 
properties of these systems \citep{deCarvalho99}. 
The one case of an X-ray group with only spiral galaxies is HCG\,16, 
whose X-ray morphology appears clumpy \citep{Dossantos99}, although this is 
not evident in recent XMM-Newton data \citep{Belsole03}.    

 Out of 69 CGs, we find 33 E-S0 dominated, and 30 Spiral dominated (6 CGs 
present a dominant member whose morphological type could not be assigned). 
For comparison, among HCGs (between 2500 and 5500 km\,s$^{-1}$) 
there are 10 Spiral dominated and 8 E-S0 dominated groups. 
The morphological content of CGs with a dominant 
E-S0 (Spiral) is listed in Table 3, along with the morphological 
content of their Neighbours.  In Table 4, we list the best fit STY 
parameters for these sample.  
\begin{table*}
\begin{center}
\caption[] {Schechter best fit parameters for CGs with (without) a dominant 
E-S0 or Spiral and for their Neighbours.}
\begin{tabular}{|l||rrc|}
\noalign{\smallskip}
\hline
\hline
\noalign{\smallskip}
sample & N$_{tot}$ & $M_{*}$ & $\alpha$ \cr 
\hline
\noalign{\smallskip}
CG (dom. E-S0 )   & 110 & --19.4  & --0.91 \\
CG (dom. Spir.)   &  92 & --19.1  & --0.72 \\
CG (without dom. E-S0)  &  77 & --18.5  & --1.02 \\
CG (without dom. Spir.) &  62 & --18.6  & --1.07 \\ 
Neigh (E-S0 dom. CGs) & 130 & --19.7  & --1.59 \\
Neigh (Spir. dom. CGs)& 129 & --20.7  & --2.10 \cr 
\noalign{\smallskip}
\hline
\hline
\end{tabular}
\end{center}
\end{table*}
The fraction of early-type galaxies is
53\% among E-S0s dominated CGs, while being 21\% among Spiral dominated CGs. 
The early-type fraction in  E-S0 dominated and Spiral dominated CGs are 
$\approx$\,typical of groups and of the field (Bahcall 1999), which could 
be used to argue that only CGs with a dominant early-type galaxy are dense and 
physical (e.g. mergers should build up the early-type fraction).    
However, when dominant galaxies are excluded from the computation 
the fraction of early-type galaxies in the two samples becomes similar 
(27\% in E-S0 dominated and 33\% in Spiral dominated).     
This could imply that densities in systems with a dominant E-S0 and a 
dominant Spiral are similar and that both type of systems are physical. 
Alternatively, assuming only E-S0 dominated CGs are physical, 
the result might suggest no global morphology-density relation to hold in 
CGs, where the morphology of the dominant galaxy alone 
traces the real underlying potential.     

The early-type fraction in samples of Neighbours of E-S0 and Spiral 
dominated CGs are similar (see Table 3) and it is noteworthy that the 
early-type fraction in both Neighbour samples tend 
to be comparable to fractions derived for the non-dominant CG population. 
This may imply that, for small groups,    
the characteristics of the brightest member are possibly a more 
fundamental parameter than general CG properties such as the total 
fraction of early-type (or late-type) galaxies, or the velocity 
dispersion \citep{Zablu,Helsdon03a}. 
Whether (and how many) CGs exist in which all members 
have their morphology modified by a locally dense environment remains an open 
question which could be checked investigating CG samples selected 
according to different criteria 
\citep{Prandoni94,Iovino02,Barton96,Giuricin00,Zandivarez03,Lee03}.  

Figure 6 shows confidence ellipses for the parameters $\alpha$ and 
$M_*$ for CGs with a dominant E-S0 (solid) and a dominant Spiral (dotted). 
The parameters are the same for E-S0s dominated and Spiral 
dominated CGs. E-S0 and Spiral dominated CGs present similar numbers of 
bright and faint galaxies and they both appear to lack faint 
galaxies relative to isolated galaxies, which display a much steeper 
faint-end LF slope.   
As expected, the exclusion of dominant members (light contours) moves  
the ellipses towards less luminous $M_*$ and steeper faint-end slopes; 
but again, no difference is found between LF parameters of Spiral 
and E-S0 dominated CGs. 

In Fig. 6 we have also plotted the contour plots of Neighbours of 
E-S0s dominated CGs (short dashed) and Spiral dominated CGs (long dashed). 
Neighbours of Spiral dominated CGs appear rich in faint and bright galaxies 
in comparison with Spiral dominated CGs. 
E-S0 dominated CGs have Neighbours which are moderately richer in faint 
and in bright galaxies. 
\begin{table*}
\begin{center}
\caption[] {Tremaine-Richstone statistic for brightest CG galaxies. }
\begin{tabular}{|l||ccccrr|}
\noalign{\smallskip}
\hline
\hline
\noalign{\smallskip}
sample &N&$\left\langle M_1 - M_2 \right\rangle$&$\sigma$($M_1$)&$t_1$&$t_1$$^{random}$&P($t_1$$^{random}$$<$$t_1$) \\\hline
\noalign{\smallskip}
CGs (all)            & 69 &0.71&0.60& 0.85  &0.90$\pm$0.10&32.3\% \\ 
CGs E-S0. dom.       & 33 &0.78&0.52& 0.67  &0.90$\pm$0.11&0.6\% \\
CGs Spir. dom.       & 30 &0.72&0.57& 0.80  &0.88$\pm$0.11&20.6\% \cr
\noalign{\smallskip}
\hline
\hline
\end{tabular}
\end{center}
\end{table*}
\begin{figure}
\resizebox{\hsize}{!}{\includegraphics{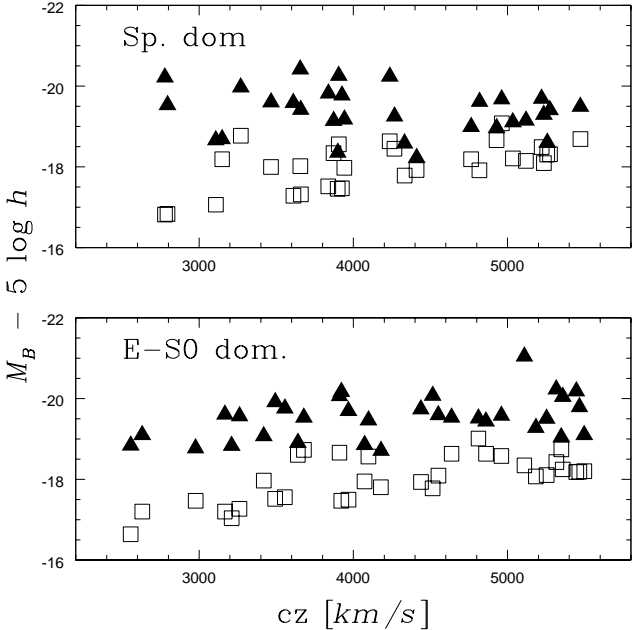}}
\hfill
\caption[]{Absolute magnitude of the brightest (triangles) 
and faintest (empty squares) member of each CG as a function of 
radial velocity for 
CGs whose dominant galaxy is an E-S0 (upper panel) and a Spiral 
(lower panels) respectively.  
}
\end{figure}
The presence of many luminous (massive) galaxies among CG Neighbours 
might indicate that dynamical friction has not operated efficiently in 
bringing bright galaxies into the center of a large group. Or, that 
bright Neighbours have an initial position so distant from the group 
centre that they had not enough time to decay by dynamical friction to the 
core of the group.  

Figure 5 and 6 indicate that CGs are poor in dwarfs relative to their 
Neighbours when applying a morphological segregation as well as 
when separating samples according to the morphological type of the CG 
dominant  galaxy. More significative differences between CGs and their 
Neighbours emerge when the comparison involves spirals or Spiral dominated 
CGs, but spirals are more numerous in both samples and small number 
statistics might affect results concerning the early-type galaxy population. 
\section{Spiral and E-S0 dominated CGs}
Both bright spirals and bright E-S0s are found in 
optically selected CGs. This agrees with the results by 
Norberg et al. (2002) indicating that (below $L_*$) both early and late 
types have approximately the same dependence of clustering strength 
on luminosity and with the finding by Cappi et al. (2003) that (very) 
bright spirals are often the brightest members of systems which 
escape standard group finding methods.  
Bright E-S0s and bright spirals in groups actually  
display different X-ray emission properties and we therefore next investigate  
whether an analogous difference  
might emerge from the analysis of optical data alone. 

We have applied the Tremaine-Richstone statistic (1977) to see if 
the brightest CG members are indeed anomalous members.  
The test is based on two parameters $t_1$ and $t_2$ defined by 
\[
 t_1 = {\sigma(M_1)  \over \langle M_2-M_1\rangle}
\]
\[
 t_2 = {\sigma(M_2-M_1)  \over (0.677)^{1/2}\langle M_2-M_1\rangle}
\]
where $\sigma$($M_1$) and $\sigma$($M_2 - M_1$) are the standard 
deviations of the absolute magnitude $M_1$ and of the difference in 
absolute magnitude ($M_2 - M_1$). Values of $t_1$ and $t_2$ below 1 
indicate that the first-ranked group galaxies are abnormally bright.  
Table 5 lists the value for $t_1$  (which appears to be a better estimator 
than the parameter $t_2$) for the whole 
sample of brightest CG galaxies and for the E-S0 and spiral dominated 
subsamples. 
Given the small number of galaxies in the CGs and the small number 
of CGs in the sample the T-R statistic $t_1$ is biased to low values 
\citep{Mamon87}. Monte-Carlo simulations (1000 random trials) 
for a given slope of the Schechter luminosity function show that the 
expected value of $t_1$ is $\simeq$\,0.90 in all samples (Mamon, priv. comm.). 
So the distribution of  ($M_1 - M_2$) from the 
galaxies in the E-S0 dominated CGs has less than 1\% probability of being 
randomly drawn from a parent LF, 
implying that the brightest group member is abnormally luminous. 
On the contrary, the dominant Spirals are not abnormally luminous. 
This is just a tentative result that should be checked on larger and 
complete CG samples. We stress that the significantly low $t_1$ for 
the E-S0 sample is partially caused by a fairly low (but not significant) 
$\sigma$($M_1$). 

Bright spirals in CGs appear therefore different compared to bright 
E-S0s in CGs. Menon \& Hickson (1985) were the first to point out that 
dominant elliptical galaxies in HCGs were special: indeed they found that 
among HCG galaxies that were radio continuum emitters, those that were 
ellipticals were always the dominant group member, 
while those that were spirals had random group rank.  
Dominant Spirals and dominant E-S0s possibly show  
a different behaviour also in the 
($m_1$ - $m_f$) parameter ($\Delta$mag between the 
brightest and the faintest CG galaxy) distribution. 
Figure 7 shows the absolute magnitude of the brightest (triangles) and the 
faintest (squares) member of each CG, plotted against radial velocity for 
for CGs with a dominant E-S0 and Spiral respectively.  

Bright Spiral dominant galaxies are preferentially hosted in nearby CGs, 
a trend which bright dominant E-S0s do not seem to follow. 
The statistical significance of the different distribution of 
($M_B$$\leq$$-19.5$) dominant E-S0s and dominant Spirals in this diagram 
emerges when comparing the different fractions of Spiral dominant galaxies 
in the cz$\leq$4000 and the cz$>$4000 km\,s$^{-1}$ subsamples. 
Below 4000 km\,s$^{-1}$ 17 bright dominant galaxies are seen, with 8 E-S0s 
and 9 Spirals.  Above 4000 km\,s$^{-1}$ 16 bright dominant galaxies are seen, 
12 E-S0s and 4 Spirals.
The most nearby dominant Spirals typically present a larger gap between 
the brightest and the faintest CG member than dominant E-S0s.   

The MW+LMC+SMC system is actually compact, and it would pass the flux 
limit if SMC did (which for $M_B$(SMC)\,=\,$-16$ translates to cz$<$2000 
km\,s$^{-1}$).    
Any CG selection criterion imposing an upper limit on the 
difference in magnitude between the brightest and the faintest group galaxy 
will certainly be biased against systems including bright spirals. 
And indeed, of the 10 dominant Spirals  
in Hickson's sample (in the 2500-5500 km\,s$^{-1}$ range) only one (HCG 90) 
is below 4000 km\,s$^{-1}$, while 4 (out of 8) dominant E-S0s are. 

Figure 7 also shows that faint-galaxy-only CGs are rare. 
Bright galaxies ($M_B$$\leq$$\sim$$-19$) 
are nearly always included in CGs whatever the magnitude of the faintest 
galaxies seen (ranging from $\simeq$$-17$ at cz=3000 km\,s$^{-1}$ 
to $\simeq$$-18$ at cz=5000 km\,s$^{-1}$).  
\section { First-ranked Spirals and E-S0s: differences in large scale 
properties}
We have shown that in UZC-CGs, dominant E-S0s are possibly anomalous 
CG members, while spirals do not share this characteristic. 
We have also shown that Spiral dominated CGs present a 
neighbourhood rich in luminous galaxies while in E-S0 dominated CGs and their 
Neighbour sample the values of $M_*$ are similar.  
These trends suggest that E-S0 dominant galaxies are more  
likely than dominant Spirals to be more luminous than their Neighbours. 
To test this hypothesis 
we next compare each CG with its own Neighbours.
We find that among the 30 dominant Spirals, 14 ($\approx$50\%) display 
a more luminous neighbour, while among the 33 dominant early-type galaxies,    
only 8 (25\%) do so.  
It is noteworthy that these 8 galaxies all are S0s, whose more luminous 
companions are generally spirals.  
We thus find that all 14 dominant Ellipticals in the CG sample are 
the (optically) brightest sources and possibly the center of the 
potential well, in a region 2\,$h^{-1}$ Mpc across. 
 
This indicates that dominant Ellipticals in CGs are typically 
the dominant member of a much larger system,  while the same is true for 
only roughly half of the dominant Spirals and S0s. That CGs and loose 
groups with a dominant Elliptical and diffuse X-ray emission 
display similar X-ray properties \citep{Helsdon00} can be used to 
indirectly support our result. 
That only half of the S0s are dominant galaxies in large groups might 
be attributed to morphological misclassification \citep{Andreon98}, 
or could imply that the evolution of Ellipticals and S0s 
is differently linked to their large scale properties. 

Concerning dominant Spirals with a more luminous companion on large scale 
(in principle similar to MW+M31), their potential well is possibly  
rather shallow, which could justify why they lack  
a diffuse X-ray halo \citep{Mulchaey03}.   
HI observations (at the GMRT) will tell whether the 16 dominant Spirals 
with no more luminous companion present 
a large unperturbed cold gas disk rather than a truncated one, 
thereby  discriminating between projected and interacting spirals.  
\begin{figure}
\resizebox{\hsize}{!}{\includegraphics{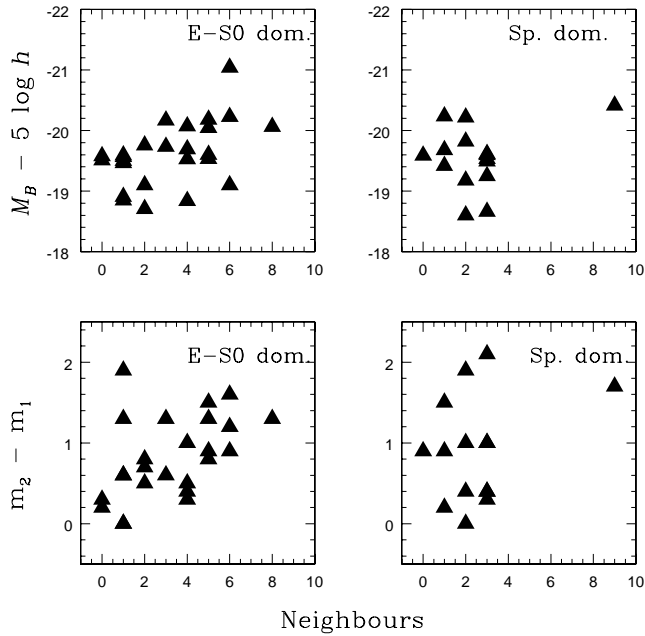}}
\hfill
\caption[]{Absolute magnitude of CG dominant galaxies (upper panels) 
as a function of the large scale density (number of Neighbours) 
for E-S0s (left) and Spirals (right) with no more luminous neighbour. 
The lower panel show the difference in magnitude ($m_2 - m_1$) 
between the first and the second ranked galaxy vs. large scale density.    
}
\end{figure} 

In Fig. 8, we show another result which is consistent 
with the picture in which E-S0 dominant galaxies 
are formed in denser and more massive groups, while 
dominant spirals are formed in lower mass, less dense environments. 
Figure 8 shows that the 25 E-S0s displaying no more luminous neighbour 
present a correlation between the density of the large scale  
environment and both, the luminosity of the galaxy and the difference  
between the magnitude of the first and second ranked galaxy.
{Spearman-rank test coefficients indicate  
that correlations exist for E-S0s ($\rho$=0.50 
in M$_B$ vs. Neighbours and $\rho$=0.51 in $\Delta$$mag_{12}$ vs. Neighbours) 
but not for spirals 
(0.05 and 0.17). For E-S0s, the correlations are 99\% significant.}

This suggests that the formation/evolution of dominant E-S0s  
and the properties of their large scale environment are possibly linked. 
Luminous passive galaxies in the 2dF \citep{Colless03} 
have already been shown \citep{Kelm03b} to display an excess of large scale 
neighbours compared to luminous emission-line galaxies. 
The location of CG dominant E-S0s at the centre of their systems to a much 
greater extent than implied by our CG selection criteria is in agreement 
with observations in the X-ray domain, showing that 
the center of diffuse X-ray emission is nearly always overlapping 
the optically brightest (early-type) galaxy of an underlying group.   
The relation between the large scale galaxy density and the difference  
between the first and second ranked galaxy could further indicate 
that, in accordance with predictions, dynamical friction has more 
efficiently operated in systems with a larger potential, 
where the fraction of group mass associated with individual galaxies is 
low \citep{Zablu}. 
This tentative result possibly extends to groups Sandage's (1976) conclusion 
that the luminosity of brightest cluster galaxies is determined by some 
special processes. Whether the formation of a group dominant 
Elliptical occurs during the formation of the group potential or later, when 
progenitor galaxies merge, is still an open question.

No relation exists between the number of CG Neighbours and 
the luminosity of CG dominant Spirals, which might  
suggest that luminous spirals in CGs are the dominant member of a small 
system whose mass is  mainly associated with the dominant member itself 
rather then with a massive halo. Therefore it seems rather unlikely 
for these systems to evolve directly into an E-S0 dominated CG.  
The alternative scenario predicts luminous spirals in CGs are contaminated 
by accordant redshift projections. 
The search within $\pm$1000 km\,s$^{-1}$ from the CG center translate to a 
search within $\pm$10\,$h^{-1}$\,Mpc, so that truly isolated galaxies, 
aligned along the line of sight of the CG will end up in the 'wrong' sample. 
Contamination is stronger in a filamentary universe 
\citep{Hernquist95}. 
Contamination of Spiral dominated CGs by Isolated galaxies 
could explain why HCGs and UZC-CGs, which should be triggered by 
interactions, do not display enhanced IR temperature nor   
IR luminosity \citep{Zepf93,Verdes-Montenegro,Kelm03a}.    
\section{Conclusions}
The morphological content and the luminosity of CGs and 
Isolated galaxies have been compared. Galaxies in the samples are 
selected from the same flux limited ($m_B$$\leq$15.5) catalogue and within the 
same redshift range (2500-5500 km\,s$^{-1}$). 
We find that galaxies in CGs display an excess of early-type galaxies and 
a lack of faint galaxies compared to isolated galaxies. 
It is mainly the spiral CG population that appears poor in faint 
galaxies suggesting that the lack of dwarf galaxies in CGs relative to 
Isolated galaxies might be related to their high content in ellipticals.

We have also compared CG galaxies with their Neighbours 
to explore whether CGs are compatible with 
being real condensations rather than temporary non physical projections 
within loose groups. 
CGs include more early-type galaxies than their Neighbours 
and they also have fewer low luminosity and high luminosity 
galaxies.  The lack in dwarfs also emerges 
when E-S0s in CGs are compared with E-S0s in the Neighbour sample 
indicating that the lack of dwarfs in CGs is not solely induced by a 
high content in E-S0. The lack of bright galaxies in CGs in comparison 
with their Neighbours is due to the presence of many bright spirals 
among Neighbours. 

In our sample, Spirals and E-S0s are equally likely to be first-ranked 
CG galaxies. It is interesting that the fraction of early type galaxies in 
E-S0 dominated and Spiral dominated CGs tends to become the same ($\sim$30\%) 
when CG dominant galaxies are excluded from computation. 

Comparing the 33 CGs with an E-S0 dominant galaxy with their 
Neighbours (and the 30 CG with a Spiral dominant galaxy with their 
Neighbours) confirms a lack of low-luminosity galaxies in CGs. 
Spiral dominated CGs  
appear to be deficient in bright members relative to their Neighbours and 
the Tremaine-Richstone test indicates that dominant Spirals 
are not anomalous CG members. E-S0s in optically selected CGs are 
however anomalous luminous members, which might relate  with the 
observation that the brightest member in X-ray emitting groups is an 
elliptical and that this elliptical is special \citep{Helsdon01,Helsdon03b}.       

All 14 dominant Ellipticals in CGs are the brightest galaxies in a region 
of redshift space 2\,$h^{-1}$\,Mpc across, while the same is true for only 
half of the dominant Spirals and dominant S0s. 
When considering only CGs whose dominant member is the brightest galaxy 
in a 1\,$h^{-1}$\,Mpc projected radius region (25 E-S0s and 16 Spirals) 
we also find a relation to link 
the number of neighbours of dominant E-S0 to 1) the luminosity of the galaxy  
and 2) the difference in magnitude between the first and second 
ranked galaxy. No such relation holds for dominant spirals. 

First ranked ellipticals appear to gain luminosity as they gain 
large scale companions implying a direct causal relationship between the 
group formation process and the formation of the first ranked galaxy therein. 
No relation exists between the luminosity of first ranked spirals 
and the number of their large scale neighbours suggesting 
that the formation of a bright spiral at the center of a large potential well 
is unlikely. 
The presence of many neighbours around dominant ellipticals is possibly  
accompanied by large quantities of infalling gas, implying that  
X-ray emission is likely associated to elliptical-dominated 
groups \citep{Zablu,Mulchaey03}.
 
In summary a clear distinction appears between CGs with a dominant 
E-S0 and CGs with a dominant Spiral. 
The former closely resemble small clusters. The latter  
are likely systems  of 1 giant + several faint galaxies, often 
presenting a bright companion within a 1\,$h^{-1}$\,Mpc distance, 
a configuration  remindful of our Local Group.  
The above results are based on a small sample, and are just tentative. 
But they all support a scenario in which the formation of   
bright Ellipticals (but not of bright spirals) and the presence of large 
group-scale potentials are linked.  
 And dominant CG Ellipticals could just represent a secondary outcome 
during the process of formation of group size systems.    
\begin{acknowledgements}
We are pleased to thank S.\,Bardelli, R.\,de\,Carvalho, 
A.\,Iovino, H.J.\,Mc\,Cracken and G.G.C.\,Palumbo, 
for stimulating discussions and suggestions. We are grateful to 
E.\,Zucca and to G. Mamon for providing scientific and technical 
support in the computation of the luminosity functions.    
The suggestions of the referee G. Mamon have been essential and have greatly 
improved the scientific content of this paper.   
This work was supported by MIUR. B.K acknowledges financial support 
from the University of Bologna. 
\end {acknowledgements}
{}
\end{document}